# Integration of Relational and Graph Databases Functionally

Jaroslav Pokorný

MFF UK, Malostranské nám. 25, 118 00 Praha 1
pokorny@ksi.mff.cuni.cz

**Abstract.** A significant category of NoSQL approaches is known as graph databases. They are usually represented by one property graph. We introduce a functional approach to modelling relations and property graphs. Single-valued and multivalued functions will be sufficient in this case. Then, a typed λ-calculus, i.e., the language of λ-terms, will be used as a data manipulation language. Some integration options at the query language level are discussed.

## 1  Introduction

A *graph database* (GDB) is based on graph theory. It uses nodes, properties, and (directed) edges [7]. A node represents an entity, such as a User, Movie, Object, and an edge represents the relationship between two nodes, e.g., is_friend_of, who_buys_what, etc. Labelled nodes and edges may have various properties (attributes) given by pairs key-value attached to them. There may be more edges between two nodes. The *property* (*attribute*) *graph* is then a multigraph. Here we suppose GDB represented by one property graph. *Graph DBMS*s (GDBMSs) proved to be very effective and suitable for many data handling use cases where relationships have a significant role. Graph querying is a key issue in any graph-based applications.

A particular case of integration of relational and NoSQL databases concerns GDBs. Today, NoSQL databases are considered in contrast to traditional RDBMS products. On the other hand, yet other approaches are possible, e.g., a functional approach. In the late 80s, there was the functional language DAPLEX [8]. In the current era of GDBMSs, we can mention Gremlin[1] - a functional graph query language developed by Apache TinkerPop which allows to express complex graph traversals. A number of significant works using functional approach to data management are contained in [2]. We will use a functional approach in which a property graph is represented by typed partial functions. We are inspired by the HIT Database Model, see, e.g., [3], as a functional alternative variant of E-R model. The functions considered will be of two kinds: single-valued and multivalued. Then, a typed λ-calculus, i.e., the language of λ-terms, can be used as a data manipulation language.

---

[1] https://tinkerpop.apache.org/gremlin.html

The aim of the paper is to introduce a functional approach to modelling relations and property graphs. In Section 2, we introduce it including querying based on a typed λ-calculus. Its usability for RDB and GDB integration is discussed in Section 3. Section 4 gives the conclusion and some proposals for future work.

## 2 Functional approach to data modelling

The functional model used here is based on a typing system. We will use elementary types and two structured types. Typed functions appropriate to modelling real data objects are *attributes* viewed as empirical typed functions that are described by an expression of a natural language [3]. The approach was studied mainly in nineties in context of conceptual modelling of databases (see, e.g., [4]).

We assume the existence of some (*elementary*) *types* $S_1,...,S_k$ ($k≥1$) constituting a *base* **B**. More complex types are constructed in the following way.

If $S, R_1,...,R_n$ ($n≥1$) are types, then
    (i) ($S:R_1,...,R_n$) is a (*functional*) *type*,
    (ii) ($R_1,...,R_n$) is a (*tuple*) *type*.

The set of *types* **T** *over* **B** is the least set containing all types from **B** and those given by (i)-(ii). When $S_i$ in **B** are interpreted as non-empty sets, then ($S:R_1,...,R_n$) denotes the set of all (total or partial) functions from $R_1×...×R_n$ into $S$, ($R_1,...,R_n$) denotes the Cartesian product $R_1×...×R_n$. Elementary type *Bool* = {TRUE, FALSE} is also in **B**. It allows to model sets (resp. relationships) as unary (resp. n-ary) characteristic functions. Logical connectives, quantifiers and predicates are typed functions, e.g., **and**/(*Bool*: *Bool*, *Bool*), quantifiers as well. Arithmetic operations are (*Number*: *Number*, *Number*)-objects, COUNT$_S$/((*Number*:(*Bool*:*S*)) is an aggregation function.

Consider **B** = {*User*, *Movie*, *U_ID*, *Name*, *Birth_y*,…}. Then, e.g., the expression "the movies rated by a user" denotes a ((*Bool*:*Movie*):*User*)-object, i.e. a (partial) function *f*:*User*→(*Bool*:*Movie*). Such (named) functions represent attributes. Each base **B** consists of descriptive and entity types. For GDBs, we can conceive entity types as sets of node IDs. Types *String*, *Number*, etc., serve for domains of properties.

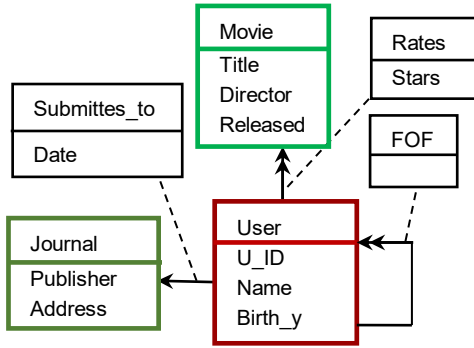

Movie/((*Title*, *Director*, *Released*):*Movie*)
User/((*U_ID*, *Name*, *Birth_y*):*User*)
Journal/((*Address*, *Publisher*):*Journal*)
FOF/((*Bool*:*User*):*User*)
Rates/((*Bool*:*Stars*, *Movie*):*User*)
Submittes_to/((*Date*, *Journal*):*User*)

**Fig. 1.** Database schema GDB Movies      **Fig. 2.** Functional schema GDB Movies

The notion of attribute applied in GDBs can be restricted to attributes of types (*R*:*S*) and ((*Bool*:*R*):*S*), where *R* and *S* are entity types. That is, single-valued and multivalued attributes are considered. Properties describing entity types are of types (($S_1$,...,$S_m$):*R*) ($m \geq 1$), where $S_i$ are descriptive types and *R* is an entity type. Similarly, we can express properties of edges. They are of types (($S_1$,...,$S_m$,$R_1$):$R_2$) or ((*Bool*:$S_1$,...,$S_m$,$R_1$):$R_2$). That is, each edge has *m* properties ($m \geq 0$).

Despite of the fact that NoSQL databases have not to use a database schema, our approach requires these structures. A GDB schema can be expressed again by a property graph (Fig. 1). Multivalued functions are described by double arrows. The schema expressed in the functional model is in Fig. 2.

Considering RDBs, (relational) attributes $A_i$:$D_j$ will be used as $S_i$. $S_i$ are non-empty sets of values, $S_i \neq S_j$ for $i \neq j$. Then relations are (*Bool*:$S_1$, ...,$S_n$)-objects. We consider relations Actors(<u>Name, Title</u>, Role) and Movies(<u>Title</u>, Released, Director, Genre), i.e., attributes

Actors/(*Bool*:*Name*, *Title*, *Role*),
Movies/(*Bool*:*Title*, *Released*, *Director*, *Genre*).

A manipulation language for functions is traditionally a typed λ-calculus (*language of terms* – LT) using typed variables, constants, applications of functions and λ-abstractions. If M/($R_1$,…,$R_n$), then *components* M[1],…,M[n] are also terms of respective types $R_1$,…, $R_n$. The language LT provides a powerful tool for querying graph data conceived as functions [5]. λ-abstractions are important here. The query „Find titles of movies directed by Spielberg", e.g., can be expressed by the term

$$\lambda\ t\ (\exists\ m,\ r\ \text{Movie}(m)(t, \text{'Spielberg'}, r)) \qquad (1)$$

Querying over a typed RDB reminds the domain relational calculus. For example,

$$\lambda\ n\ (\forall t\ (\exists\ re,\ g\ \text{Movies}(t, re, \text{'Spielberg'}, g)\ \textbf{implies}\ \exists\ ro\ \text{Actors}(n, t, ro))) \qquad (2)$$

expresses the query "Find the actors, who play in each film by director Spielberg."

Regardless of expressive power, LT terms are not too user-friendly. The following versions of (1) and (2), respectively, indicate how to (partially) solve this problem:

$\{t^{Title} \mid \textbf{exists}\ m^{Movie}\ \text{Movie}(m^{Movie})(t^{Title}, \text{'Spielberg'}^{Director})\}$
$\{n^{Name} \mid \textbf{foreach}\ t^{Title}\ (\text{Movies}(t^{Title}, \text{'Spielberg'}^{Director})\ \textbf{implies}\ \text{Actors}(n^{Name}, t^{Title}))\}$

## 3    Integration of relations and property graphs

The book [1] offers three ways of integration of the two different worlds of relational and NoSQL databases: *native*, *hybrid*, and *reducing to one option* (either relational or NoSQL). Several approaches are under a development [6]: *polyglot persistence*, *multi-model approach*, *multilevel modelling*, *NoSQL relationally* and *schema and data conversion*. The most relevant for our approach is a multilevel modelling covering the following subapproaches: (a) *special abstract model*, (b) *NoSQL-on-RDBMS*, or (c) *ontology integration*. We are closed to the (a) subapproach, where starting model is the functional data model. As the functional modelling provides rather hybrid between a conceptual and database schema, the c) subapproach is also relevant.

LT queries sent to the integrated system are translated into queries compatible with the RDBMS (e.g., SQL) and GDBMS (e.g., Cypher language of Neo4j), respectively.

On the data level, associated databases are generally heterogeneous. Elementary type *Title* of movies has not to be the same as the domain(*Title*) from the Movies relation. Only their non-empty intersection should be supposed. To integrate both schemas, we used a renaming some attributes and/or relations (e.g. Movie vs. Movies). Then, the term

$$\lambda\, u^{User},\, g^{Genre},\, n^{Number}\, (n^{Number} = \text{COUNT}_{Movie}\, (\lambda\, m^{Movie}\, (\text{Rates}(u^{User})(m^{Movie})\, \textbf{and}$$
$$\exists\, t^{Title}\, s^{Title}\, \text{Movie}(m^{Movie}).t^{Title} = s^{Title}\, \textbf{and}\, \text{Movies}(s^{Title},\, g^{Genre}))\,))$$

expresses the query, "Find for each user and genre the number of reviews done by him/her in the genre".

## 4 Conclusions

In the paper, we have focused on GDBs based on property graphs as a NoSQL data source and RDBs. Current challenges for database research of such infrastructure include:
- finding an appropriate and successfully powerful subset of LT covering query requirements of a GDB integrated with an RDB,
- developing a meaningful and usable user-friendly version of a query language based on LT,
- developing a prototype using an SQL engine and Neo4j GDBMS for source databases.

**Acknowledgments.** This work was supported by the Charles University project Q48.